# Global Magnitude of the Earthquakes

A.V. Guglielmi [a)],  B.I. Klain [b)]

[a)] *Institute of Physics of the Earth RAS, Moscow, Russia,*
*guglielmi@mail.ru*

[b)] *Borok Geophysical Observatory of IPE RAS, Yaroslavl Region, Borok, Russia,*
*klb314@mail.ru*

**Abstract**

The idea of the global daily magnitude (GDM, unit designation is Mg) of earthquakes is introduced. The formula is given for calculating Mg from earthquake strength data, characterized by the classic magnitude M on the Richter scale. An example of the calculation of Mg is given. It is noted that the Mg index may be useful in studying a number of seismological problems.

## 1. Introduction

It is well known [1] that the energy E of a particular earthquake in the "ideal" energy scale is connected by a certain ratio with the magnitude M on the Richter scale:

$$\lg E = a + b M . \qquad (1)$$

Here $a = 4.8$, and $b = 1.5$ if E is expressed in joules. In this short paper we propose introducing the global daily magnitude (GDM) as a convenient characteristic of earthquakes that occurred over a calendar day. GDM may be considered as a useful index of the planetary earthquakes activity in various studies in the seismology (see Section 3). It is quite clear that when characterizing global seismicity, one cannot directly use the Richter magnitude, since it is not an additive quantity. The energy is additive, and on this basis we propose the following formula for calculating global magnitude.

$$\mathrm{Mg} = \frac{1}{\beta} \ln\left[\sum_j H_j \exp(\beta M_j) \Big/ \sum_j H_j\right]. \qquad (2)$$

Here $\beta = b\ln 10$, and $H_j$ is the dichotomous variable (the Heaviside symbol) taking the values 0 if $M_j < M_0$, and 1 if $M_j \geq M_0$. The choice of magnitude $M_0$ is dictated by the research task and the properties of the global earthquake catalog, which is used to calculate the Mg. Index $j$ = 1, 2, 3, … runs through a segment of the natural series of numbers. It lists the earthquakes that occurred during the day and registered in the catalog.

## 2. Example

Here is an example of calculating GDM using the formula (2). We use the USGS catalog (https://earthquake.usgs.gov/earthquakes/). We have chosen the parameter $M_0$ being equal to 1.

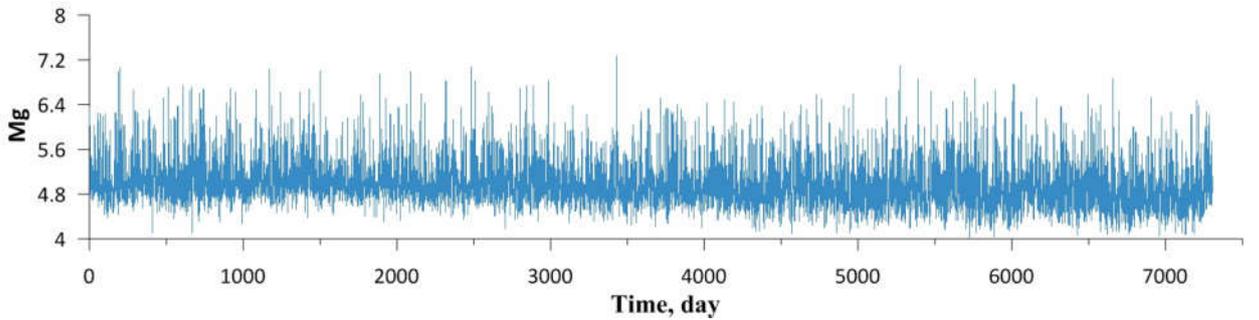

Fig. 1. The series of GDM values from 01/01/1980 to 12/31/1999.

Figure 1 gives us an idea of the GDM variations for 20 years from 01/01/1980 to 12/31/1999. When $M_0 = 1$ is selected in formula (2), the Mg value changes from 3.8 to 7.3.

Figure 2 shows the distribution of events by the global daily magnitude Mg. We see that the representative part of the distribution obeys the Gutenberg-Richter law:

$$\lg N = 8.91 - 1.13 \mathrm{Mg}. \qquad (3)$$

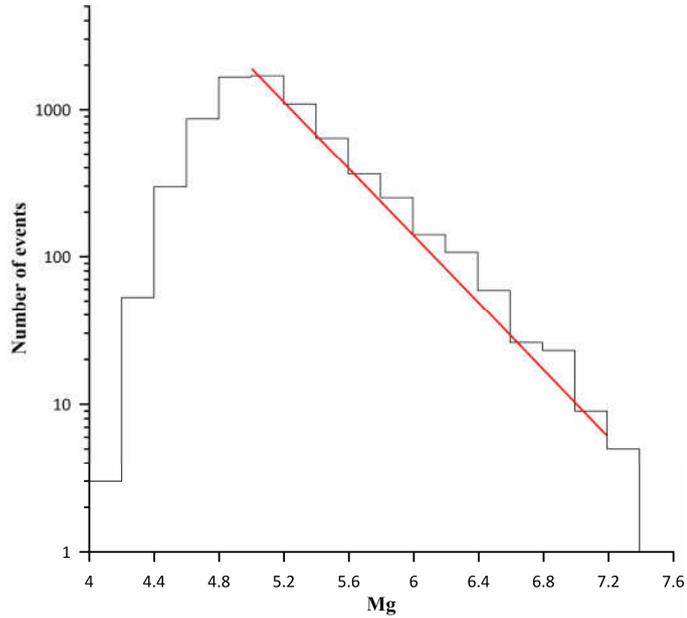

Fig. 2. Distribution of the events for the Mg value.

## 3. Discussion

### a. GDM peak statistics

Let us discuss the possible applications of GDM. First of all, the significant variability of the Mg series shown in Figure 1 is of interest. We see a slow trend with a general decrease in GDM by the end of a 20-year period. Apparently, it reflects the secular variations of the Earth's seismic activity. Of particular interest are rare spikes formed by abrupt events with high Mg values. Let's take a closer look at them.

Looking at Figure 3, we can assume that the distribution of GDM peaks obeys the Poisson statistics

$$p(k) = \frac{\lambda^k}{k!}\exp(-\lambda), \tag{4}$$

with the parameter $\lambda = 3.1$. Here $\lambda$ is the average number of events over a fixed period (over 100 days in our case).

However, there is one complicating circumstance. Namely, our series of random strong events ($Mg \geq 6$) is not stationary. The value of $\lambda$ varies from 3.8 at the beginning to 2.3 at the end of a 20-year period. We assume that variation of $\lambda$ shows once again the secular changes of seismicity. Obviously, the question of the long-term variation of the statistical properties of strong events deserves further study.

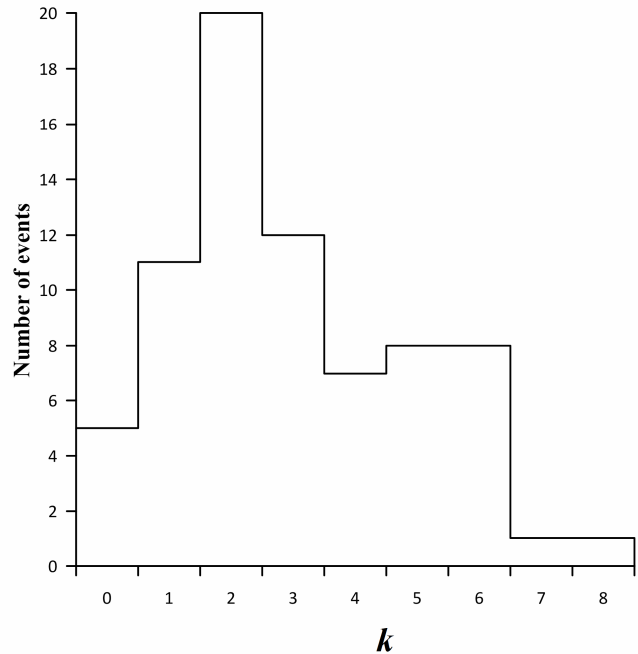

Fig. 3. Distribution of rare events with $Mg \geq 6$. Here $k$ is the number of events over the time interval of 100 days.

**b. GDM and Wolf number**

Our work has been stimulated by the idea of the magnetoplasticity of rocks presented in the papers [2,3]. Briefly, the idea is that under the influence of an alternating magnetic field of artificial or natural origin, the plasticity of the rocks composing the earth's crust increases, which leads to a noticeable change in the regime of seismic activity (see the cited papers for more details).

We intend to study the relationship of Mg variations with various manifestations of geomagnetic activity. In this paper, we restrict ourselves to a simpler problem. Namely, we compare the seismic activity characterized by the Mg index with the Wolf numbers W, which reflect the solar activity. (For a relation between solar and geomagnetic activity see for example [4].)

The points in Figure 4 give a general idea that the dependence of Mg on W, if it exists, firstly is weak, and secondly rather complicated. We will make one argument in favor of the assumption that this dependence nevertheless exists in reality.

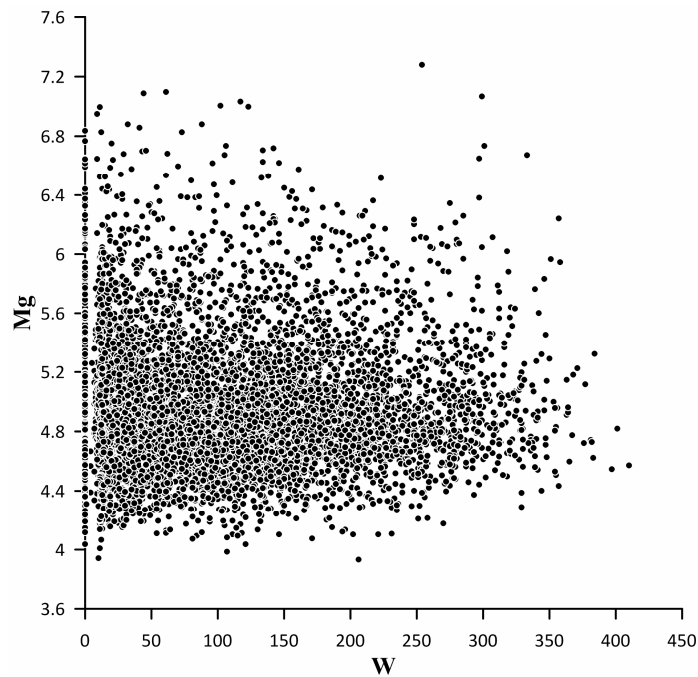

Fig. 4. Scatterplot reflecting the possible statistical relationship between Mg and W.

From the set of all events, we distinguished two subsets, namely, the lower and upper sextiles, containing events with relatively small and large Mg values, respectively. Each of the two subsets contains 1217 pairs of W and Mg values. Next, we calculated the average values of W for each subset: $W = 93.6 \pm 2$ and $W = 104.1 \pm 2$ (see Figure 5). The difference between Wolf averages is 10.5. The well-known rule of "three sigma" is fulfilled with a margin. Thus, a certain dependence of Mg on W exists reliably at a rather high level of statistical significance.

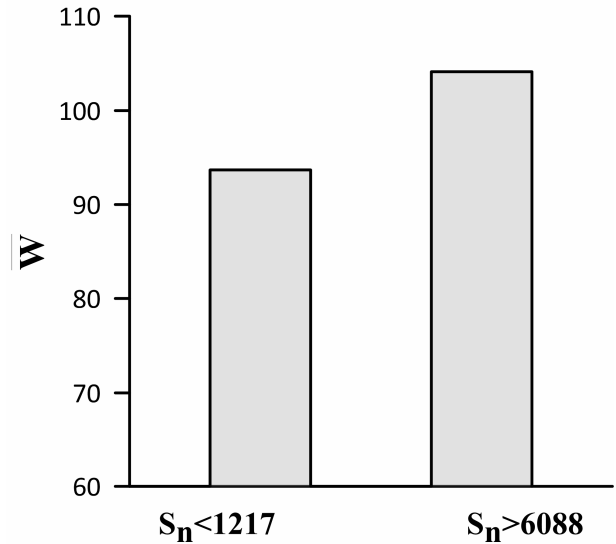

Fig. 5. Average Wolf numbers for the days with a weak and strong seismic activity (the left and right columns respectively, see the text).

**c. Modifications of GDM**

When calculating the Mg in the present study, we chose a time interval equal to the calendar day. However, nothing prevents us from choosing another interval, if this is dictated by the task of research. For example, we can choose a three-hour interval when studying the effects on the planetary seismicity of geomagnetic activity, characterized by the Ap index. The hourly interval is also convenient if we are interested in the relationship between earthquakes and geomagnetic storms, which are characterized by the Dst index. On the contrary, a larger interval, for example, annual, is useful in the study of secular variations of the Earth's seismicity. Special attention should be the selection of parameter $M_0$. Our choice of $M_0 = 1$ in the illustrative example in Figure 1 was made only for reasons of convenience. In a more rigorous approach, generally speaking, one should use the representative part of the catalog and, in accordance with this, select the $M_0$ value.

Finally, it is worth mentioning the possibility of using a formula similar to (2) to calculate the regional magnitude Mr. In this case, it may be useful to use the Heaviside symbol to highlight not relatively strong, but weak earthquakes. The point here is as follows. In the work [5] the so-called weekend effect is described, which manifests itself namely in the activity of relatively weak earthquakes.

**4. Conclusion**

We have presented the formula (2) to calculate the global magnitude Mg, characterizing the seismicity of the Earth as a whole. We have given a concrete example of the calculation of Mg. An analysis of this example showed that the Mg index may be useful in studying some problems of seismology.

*Acknowledgments*. We are grateful to A.L. Buchachenko, A.S. Potapov, A.D. Zavyalov and O.D. Zotov for numerous discussions of the problems of seismology and solar-terrestrial physics. We sincerely thank I.P. Lavrov for helping with the earthquake catalog. Results were partially obtained using the USGS catalog (https://earthquake.usgs.gov/earthquakes/). The work was supported by the Program No. 12 of the RAS Presidium, project of the Ministry of Education and Science of the Russian Federation KP19-270, project of the RFBR 18-05-00096, as well as the state assignment program of the IPhE RAS.